# J-matrix method of scattering in any $L^2$ basis

**H. A. Yamani[1], A. D. Alhaidari[2], and M. S. Abdelmonem[2]**
[1]*Ministry of Industry & Electricity, P. O. Box 5729, Riyadh 11127, Saudi Arabia*
E-mail: haydara@sbm.net.sa
[2]*Physics Department, King Fahd University of Petroleum& Minerals, Box 5047, Dhahran 31261, Saudi Arabia*
E-mail: haidari@mailaps.org, msmonem@kfupm.edu.sa

The restriction imposed on the J-matrix method of using specific $L^2$ bases is lifted without compromising any of the advantages that it offers. This opens the door to a wider range of application of the method to physical problems beyond the restrictive SO(2,1) dynamical symmetry. The numerical scheme developed to achieve this objective projects the J-matrix formalism in terms of the eigenvalues of a finite Hamiltonian matrix and its submatrices in any convenient $L^2$ basis. Numerical stability and convergence of the original analytic J-matrix method is still maintained in the proposed scheme, which can be applied to multi-channel nonrelativistic as well as relativistic scattering problems.

## I. INTRODUCTION

The utilization of $L^2$ methods to carry out scattering calculation underwent several major developments starting with the stabilization technique of Hazi and Taylor [1]. Then followed by the development of the $L^2$-Fredholm method [2] in which scattering information is extracted from the discrete eigenvalues of the scattering Hamiltonian, $\{\varepsilon_n\}_{n=0}^{N-1}$, and those of the reference Hamiltonian, $\{\varepsilon_n^0\}_{n=0}^{N-1}$, in a finite $L^2$-basis, $\{|\chi_n\rangle\}_{n=0}^{N-1}$. This method is analogous to discretizing the continuous spectrum of the system by confining it to a box in configuration space. Although the approximate Fredholm determinant, $\tilde{D}(z)$, as given by

$$\tilde{D}(z)=\prod_{n=0}^{N-1}\left(\frac{z-\varepsilon_n}{z-\varepsilon_n^0}\right) \tag{1.1}$$

does not have the analytic structure of the exact Fredholm determinant, $D(z)$, Reinhardt and his group [3] postulated that $\tilde{D}(z)$ is a good approximation to $D(z)$ in the complex plane away from the real energy axis. There, $\tilde{D}(z)$ can be fitted to a low order continued fraction and then analytically continued to the real energy axis with the resulting phase identified as the negative of the scattering phase shift.

Subsequent development was based on the realization that for certain $L^2$ bases the approximation (1.1) may be interpreted as a quadrature approximation to the known dispersion equation relating the Fredholm determinant to its imaginary part, $A(E)$, as

$$D(z)=1-\frac{1}{\pi}\int_0^\infty \frac{A(E)}{z-E}dE \tag{1.2}$$

These specified bases made it possible to relate the energy eigenvalues, $\{\varepsilon_n^0\}_{n=0}^{N-1}$, closely to the zeros of a known orthogonal polynomial $p_N(z)$.

Heller and Yamani [4] then realized that in some complete Laguerre and Oscillator bases of $L^2$ function space it was not necessary to confine the reference Hamiltonian, $H_0$, to a finite matrix representation as is done to the short range scattering potential. This is because these special bases allow a full algebraic solution to the $H_0$–problem. This basic concept of the J-matrix method [4-6] endowed the method with both formal and computational similarities to the R-matrix method [7]. The J-matrix method yields *exact* scattering information over a continuous range of energy for a model potential obtained by truncating the given short-range potential in a finite subset of this basis. It was shown to be free from fictitious resonances that plague some algebraic variational scattering methods [8]. The method has been applied successfully to a large number of non-relativistic problems and has been extended to multi-channel [9,10] as well as relativistic [11,12] scattering. The accuracy and convergence property of the method compares favorably with other successful scattering calculation methods [13]. Attempts have been made to account for the class of all Hamiltonians that are compatible with the J-matrix formalism [14]. Beside the familiar Coulomb, oscillator, and S-wave Morse potentials very few other potentials were added to this class which included some interesting potentials at zero energy. It turns out that SO(2,1) is an underlying symmetry of the theory [14,15] which, of course, restricts the scope of application of the method to physical problems with only this dynamical symmetry group. Not only the kind of reference Hamiltonians is restricted but so is the choice of a convenient basis. To start with, one has to find a basis in which the representation of the Schrödinger (wave) operator of the unperturbed problem is tridiagonal. This meant that both the matrix representation of the reference Hamiltonian and the basis-overlap matrix must be at most tridiagonal. Thereafter, one has to solve the resulting recursion relation to obtain the required asymptotic expansion coefficients of the regularized wave function. In some instances, even if it was possible to arrive at the required tridiagonal representation, the resulting recursion relation may turn out to be too difficult to solve analytically. This is for example the case in the relativistic Dirac-Coulomb problem [12].

Although the J-matrix method found applications in atomic and nuclear scattering, and its results compared favorably with other methods, it was not as widely used as other methods such as the R-matrix. One reason is that the method required some knowledge of more mathematical functions than that of other methods. But more importantly, the J-matrix method was restrictive in the use of the $L^2$ bases to some special bases sets in which the $H_0$–problem is analytically soluble. Other $L^2$ bases exist that are sometimes more useful than the J-matrix bases.

In this paper we lift the restriction imposed on the method of using specific $L^2$ bases. We show that any given complete set of $L^2$ bases, $\{|\chi_n\rangle\}_{n=0}^{\infty}$, can be fitted into the J-matrix method without compromising any of the advantages that it offers. Since the tridiagonal matrix representation of $H_0$ is basic to the analytic structure of the theory, then even if the given representation in the conveniently chosen basis is not tridiagonal, we can still use numerical routines (e.g. the well-known Householder method) to tridiagonalize it [16]. The problem is, therefore, transformed back into the framework of the J-matrix method enabling the use of the usual computational tools of the formalism. The tridiagonalization routines also produce an orthogonal transformation matrix that performs the change to the new basis in which $H_0$ is tridiagonal. Since all kinematic information is in $H_0$, which is not altered by the unitary transformation, then we should expect that the diagonal and off-diagonal entries in the resulting tridiagonal matrix carry this information. Specifically, we expect that these entries are all that is required to compute the J-matrix kinematical coefficients needed for the calculation of the S-matrix. We have already developed a numerical scheme based on this technique. It utilizes the continued fraction representation of the J-matrix objects in terms of the entries of the tridiagonal matrices. However, due to numerical rounding errors associated with tridiagonalization reduction routines for matrices in some bases, the accuracy of the algorithm may be compromised [16,17]. Therefore, we propose here an alternative scheme.

The proposed scheme similarly does not require a tridiagonal representation for the reference Hamiltonian or the basis overlap matrix in the given $L^2$ basis. It projects the J-matrix formalism in terms of the *eigenvalues* and *eigenvectors* of these matrices and their submatrices rather than the elements of their tridiagonal representations. Therefore, no explicit reference is made to the tridiagonal structure of the theory. In fact, the scheme solves the recursion relation indirectly giving the J-matrix kinematical coefficients needed for the calculation of the S-matrix. This has at least three major advantages. First, it opens the door to a wider range of physical problems beyond the restrictive SO(2,1) symmetry. Second, it gives the freedom of choice for any suitable and convenient basis without the constraint of a tridiagonal representation neither for the reference Hamiltonian nor for the basis-overlap matrix. Third, it eliminates the need for recursive calculations associated with recursion relations, which are usually plagued with numerical instabilities. Most importantly, this numerical scheme can be applied to non-relativistic as well as relativistic scattering problems as demonstrated in one of the examples. The foundation of the proposed scheme is based on the theory of orthogonal polynomials associated with tridiagonal matrices.

The paper is organized as follows: In section II we identify the quantities needed as computational tools in the J-matrix formalism. In section III we describe how to calculate these J-matrix quantities without reference to the recursion relation or to the tridiagonal representations. We also outline the steps of the numerical algorithm for computing the S-matrix. In section IV we demonstrate the utility, accuracy, and ease of the proposed numerical algorithm by applying it to some examples and compare that with known analytic results. In the Appendix, we give the necessary modifications to the scheme in the case of non-orthogonal basis.

## II. THE COMPUTATIONAL TOOLS OF THE J-MATRIX METHOD

Let $H$ be the total Hamiltonian of the system which we can write as

$$H = H_\infty + V \tag{2.1}$$

where $H_\infty$ is the Hamiltonian of the system when the incident particle is at an infinite separation from the target where $V = 0$. Let's assume that the potential can be written as

$$V = V_0 + \tilde{V} \tag{2.2}$$

where $\tilde{V}$ is short range and $V_0$ is the analytical part of the potential with the property that when added to $H_\infty$ gives the reference Hamiltonian

$$H_0 \equiv H_\infty + V_0 \tag{2.3}$$

which admits a tridiagonal matrix representation in a proper $L^2$ basis $\left\{|\varphi_n\rangle\right\}_{n=0}^{\infty}$:

$$H_0 = \begin{pmatrix} a_0 & b_0 & & & & & \\ b_0 & a_1 & b_1 & & \mathbf{0} & & \\ & b_1 & a_2 & b_2 & & & \\ & & b_2 & \times & \times & & \\ & & & \times & \times & \times & \\ & \mathbf{0} & & & \times & \times & \times \\ & & & & & \times & \times \end{pmatrix} \tag{2.4}$$

This means that

$$J_{nm} \equiv \langle \varphi_n | (H_0 - E) \; |\varphi_m\rangle = 0 \qquad \text{for} \quad |n-m| \geq 2 \tag{2.5}$$

We call $|\psi\rangle$ the solution of the $H_0$-problem is the sense that

$$\langle \varphi_n | (H_0 - E) |\psi\rangle = 0 \qquad ,\forall n \tag{2.6}$$

outside a dense region of space around the origin of *V*.
Expanding $|\psi\rangle$ in this basis as $|\psi\rangle = \sum_n d_n |\varphi_n\rangle$, we can write Eq. (2.6) as

$$\begin{aligned} J_{n,n-1} d_{n-1} + J_{nn} d_n + J_{n,n+1} d_{n+1} &= 0 \qquad ,n \geq 1 \\ J_{00} d_0 + J_{01} d_1 &= 0 \end{aligned} \tag{2.7}$$

Typically, there exist two independent solutions to (2.6) which behave asymptotically like free particles, one of them as $\sqrt{2/\pi k}\,\sin(kr)$ and the other as $\sqrt{2/\pi k}\,\cos(kr)$, where $k = \sqrt{2E}$. We write these two solutions, respectively, as:

$$\begin{aligned} |\psi_s(E)\rangle &= \sum_n s_n(E) |\varphi_n\rangle \\ |\psi_c(E)\rangle &= \sum_n c_n(E) |\varphi_n\rangle \end{aligned} \tag{2.8}$$

The set $\{s_n\}$ satisfy the recursion relation (2.7), whereas $\{c_n\}$ satisfy the recursion relation (2.7) only for $n \geq 1$ while the last relation reads [5]

$$J_{00} c_0 + J_{01} c_1 = k/2s_0 \tag{2.9}$$

The two sets, $\{s_n\}$ and $\{c_n\}$, are related by the Wronskian-like relation

$$J_{n,n-1}\left(c_n s_{n-1} - c_{n-1} s_n\right) = k/2 \qquad ;n \geq 1 \tag{2.10}$$

and are tabulated elsewhere [18].

Heller [8] showed that the Green's function *G*(*E*) which is formally defined by the relation

$$G(E)\, J(E) = 1 \tag{2.11}$$

has the matrix representation

$$G^{\pm}_{nm}(E) = \langle \bar{\varphi}_n | G(E \pm i0) |\bar{\varphi}_m\rangle = \frac{2}{k} s_{n_<}(E) \left[ c_{n_>}(E) \pm i s_{n_>}(E) \right] \tag{2.12}$$

where $n_>$ ($n_<$) is the larger (smaller) of the two indices *n* and *m* and $\langle \bar{\varphi}_n |$ is an element of the conjugate basis satisfying $\langle \bar{\varphi}_n | \varphi_m \rangle = \delta_{nm}$.

Now it is assumed that $\tilde{V}$ is short range and such that it can be approximated well by the potential $\tilde{V}^{(N)}$, for large enough *N*, whose matrix representation is given by

$$\tilde{V}_{nm}^{(N)} = \begin{cases} \langle \varphi_n | \tilde{V} | \varphi_m \rangle & ; n, m \le N-1 \\ 0 & \text{otherwise} \end{cases} \tag{2.13}$$

We call $|\Psi\rangle$ the solution of the $\left(H_0 + \tilde{V}^{(N)}\right)$-problem in the sense that

$$\langle \varphi_n | \left(H_0 + \tilde{V}^{(N)} - E\right) |\Psi\rangle = 0 \qquad \forall n \tag{2.14}$$

It has been shown [4-6] that the exact scattering matrix for this problem is given by

$$S(E) = T_{N-1}(E) \left[ \frac{1 + g_{N-1,N-1}(E) J_{N-1,N}(E) R_N^-(E)}{1 + g_{N-1,N-1}(E) J_{N-1,N}(E) R_N^+(E)} \right] \tag{2.15}$$

It is clear from this equation that the needed computational tools are three quantities, namely:

(i) $g_{n,m}(E) = \langle \bar{\varphi}_n | \, (H_0 + \tilde{V}^{(N)} - E)^{-1} \, | \bar{\varphi}_m \rangle$ which is the finite-matrix version of the full Green's function,

(ii) $T_n \equiv \dfrac{c_n - i s_n}{c_n + i s_n}$, and (2.16.1)

(iii) $R_{n+1}^{\pm} \equiv \dfrac{c_{n+1} \pm i s_{n+1}}{c_n \pm i s_n}$ (2.16.2)

$g_{n,m}(E)$ is computationally easy to calculate and we will comment on its evaluation in the next section. We thus concentrate our effort on how to accurately calculate $T_n$ and $R_{n+1}^{\pm}$ in the chosen $L^2$ basis.

Had the reference Hamiltonian's representation been tridiagonal, it would have been possible to use the finite continued fraction representation, developed by Yamani and Abdelmonem [10], of $T_n(E)$ and $R_{n+1}^{\pm}(E)$ in terms of the tridiagonal matrix elements $J_{nm}(E)$. Even if the matrix representation of $H_0$ in the chosen basis is not tridiagonal to start with, it is possible to numerically tridiagonalize it, and hence utilize the resulting coefficients to implement the scheme mentioned above and calculate $T_n(E)$ and $R_{n+1}^{\pm}(E)$. We, however, choose not to pursue this scheme to avoid numerical errors that inevitably result from the tridiagonalization step. We choose instead to base our calculation on a scheme that utilizes only the energy eigenvalues of $H_0$, and other related matrices, to calculate $T_n(E)$ and $R_{n+1}^{\pm}(E)$. In the next section we show how to accomplish the goal by utilizing the properties of tridiagonal matrices and associated orthogonal polynomials.

## III. $T_n(E)$ AND $R_{n+1}^{\pm}(E)$ IN TERMS OF THE EIGENVALUES OF $H_0$ AND RELATED MATRICES

It is known from the theory of orthogonal polynomials [19] that associated with any symmetric tridiagonal matrix, like $H_0$, are two sets of orthogonal polynomials $\{p_n(z)\}_{n=0}^{\infty}$ and $\{q_n(z)\}_{n=0}^{\infty}$ defined, for any complex *z*, by

$$\begin{aligned} &p_0(z) = 1 \quad \text{and} \quad p_1(z) = (z - a_0)/b_0 \\ &q_0(z) = 0 \quad \text{and} \quad q_1(z) = 1/b_0 \end{aligned} \tag{3.1}$$

and satisfying the three-term recursion relation (2.7), namely

$$zd_n(z) = a_n d_n(z) + b_{n-1} d_{n-1}(z) + b_n d_{n+1}(z) \quad ; \quad n \geq 1 \tag{3.2}$$

where $d_n(z)$ stands for either $p_n(z)$ or $q_n(z)$. Therefore, by comparison of the initial conditions and recursion relation for the polynomials $\{p_n, q_n\}$ with those of $\{s_n, c_n\}$, it is evident that we can always make the following identification

$$p_n(z) = s_n(z)/s_0(z) \quad \text{, and}$$
$$q_n(z) = \frac{2}{k}\left[c_n(z)s_0(z) - c_0(z)s_n(z)\right] \tag{3.3}$$

The polynomial $p_n(z)$ is of order $n$ in $z$ and $q_n(z)$ is a polynomial of order $(n-1)$ in $z$. The set of $n$ zeros, $\{\varepsilon_k\}_{k=0}^{n-1}$, of $p_n(z)$ are the eigenvalues of the finite $n\times n$ matrix $H_0$. The set of $n-1$ zeros, $\{\hat{\varepsilon}_k\}_{k=0}^{n-2}$, of $q_n(z)$ are the eigenvalues of the abbreviated version of this matrix obtained by deleting the first raw and first column [19]. These polynomials also satisfy the Wronskian-like relation

$$b_k\left[p_k(z)q_{k+1}(z) - p_{k+1}(z)q_k(z)\right] = 1 \quad ; \quad k \geq 0 \tag{3.4}$$

which parallels that of Eq. (2.10) for the sine-like and cosine-like solutions of the reference wave equation.

The element $G_{00}(z)$ has a particularly simple limiting representation

$$G_{00}(z) = -\lim_{n\to\infty}\left\{\frac{q_n(z)}{p_n(z)}\right\} \tag{3.5}$$

It is an analytic function in the complex $z$-plane with a branch cut on the positive real energy axis. The general matrix element $G_{nm}(z)$ is related to the element $G_{00}(z)$ and the polynomials $\{p_n, q_n\}$ as

$$G_{nm}(z) = p_n(z)p_m(z)\left\{G_{00}(z) + \frac{q_{n_>}(z)}{p_{n_>}(z)}\right\} \tag{3.6}$$

where $n_>$ is the larger of $n$ and $m$. Combining Eq. (2.12) with (2.16), we obtain

$$T_{N-1}(E) = \frac{G^-_{0,N-1}(E)}{G^+_{0,N-1}(E)} \quad \text{and} \quad R^\pm_N(E) = \frac{G^\pm_{0,N}(E)}{G^\pm_{0,N-1}(E)} \tag{3.7}$$

Using (3.6) we can write these as

$$T_{N-1}(E) = \frac{G^-_{00}(E) + [q_{N-1}(E)/p_{N-1}(E)]}{G^+_{00}(E) + [q_{N-1}(E)/p_{N-1}(E)]} \tag{3.8}$$

and $$R^\pm_N(E) = \left[\frac{p_N(E)}{p_{N-1}(E)}\right]\frac{G^\pm_{00}(E) + [q_N(E)/p_N(E)]}{G^\pm_{00}(E) + [q_{N-1}(E)/p_{N-1}(E)]} \tag{3.9}$$

The polynomial ratio $\left[q_n(z)/p_n(z)\right]$ has the following representation in terms of $\{\varepsilon_k\}_{k=0}^{n-1}$ and $\{\hat{\varepsilon}_k\}_{k=0}^{n-2}$

$$\frac{q_n(z)}{p_n(z)} = \frac{\prod_{k=0}^{n-2}(z-\hat{\varepsilon}_k)}{\prod_{k=0}^{n-1}(z-\varepsilon_k)} \tag{3.10}$$

On the other hand the polynomial ratio $\left[p_n(z)/p_{n-1}(z)\right]$ has the following representation in terms of $\{\varepsilon_k\}_{k=0}^{n-1}$ and $\{\tilde{\varepsilon}_k\}_{k=0}^{n-2}$

$$\frac{p_n(z)}{p_{n-1}(z)}=\frac{1}{b_{n-1}}\frac{\prod_{k=0}^{n-1}(z-\varepsilon_k)}{\prod_{k=0}^{n-2}(z-\tilde{\varepsilon}_k)} \tag{3.11}$$

where $\{\tilde{\varepsilon}_k\}_{k=0}^{n-2}$ are the eigenvalues of the $(n-1)\times(n-1)$ submatrix of $H_0$ obtained by deleting the last row and last column of the $n\times n$ matrix $H_0$. The element $b_{N-1}$ that enters in the polynomial ratio $\left[p_N(E)/p_{N-1}(E)\right]$ is a tridiagonalization-dependent term. However, due to the fact that $R_N^{\pm}$ enters in the S-matrix multiplied by $J_{N-1,N}$, as seen in (2.15), and since $J_{N-1,N}(E)=b_{N-1}$ in orthogonal basis, this dependence is eliminated. Thus, we see that the accurate calculation of $T_n(E)$ and $R_{n+1}^{\pm}(E)$ is related to finding an accurate and reliable scheme of calculating $G_{00}^{\pm}(E)$.

The analyticity of $G_{00}(z)$ in the complex energy plane cannot be maintained for any finite approximation of a representation such as that in Eq. (3.5) no matter how large $n$ is. Let $\tilde{G}_{00}(z)$ stand for the finite $M$-term version of (3.5). That is, given an $M\times M$ matrix $H_0$ in the conveniently chosen orthogonal basis, $\{\chi_n\}_{n=0}^{M-1}$, we write the associated finite Green's function using (3.10) as

$$\tilde{G}_{00}(z)=-\frac{\prod_{k=0}^{M-2}(z-\hat{\varepsilon}_k)}{\prod_{k=0}^{M-1}(z-\varepsilon_k)} \qquad ;M\geq N \tag{3.12}$$

It possesses a set of $M$ real poles, $\{\varepsilon_k\}_{k=0}^{M-1}$, interlaced with $M-1$ real zeros, $\{\hat{\varepsilon}_k\}_{k=0}^{M-2}$, which try to mimic the cut structure of $G_{00}(z)$ on the real energy axis. However, $\tilde{G}_{00}(z)$ is a good approximation to $G_{00}(z)$ in regions of the complex plane away from the real axis. Thus the evaluation of $\tilde{G}_{00}(z)$ at a set of $K$ points $\{z_i\}_{i=0}^{K-1}$ in the upper half of the complex plane gives complex values that can be analytically continued back to the real axis giving an approximation to $G_{00}(E)$. The set of complex points has to be chosen far enough away from the real axis so as to smooth the effect of the pole structure of $\tilde{G}_{00}(z)$. Yet they should be close enough to the real axis so that the information contained in $\tilde{G}_{00}(z)$ is not washed out in the continuation process [20,21]. The set $\left\{z_i,\tilde{G}_{00}(z_i)\right\}_{i=0}^{K-1}$ is fitted to a low order rational fraction approximation $F(z)$ of Schlessinger-Haymaker [22]. $F(z)$ provides a smooth analytic continuation back to the real axis when taking the limit $z\to E+i0$.

We thus summarize the steps needed to carry out the J-matrix calculation in the chosen orthogonal basis $\{\chi_n\}$:

1. We calculate $\left\{\tilde{V}_{nm}\right\}_{n,m=0}^{N-1}$ either analytically or numerically depending on the nature of the basis $\{\chi_n\}_{n=0}^{\infty}$.

2. The matrix elements of $H_0$, in the basis $\{\chi_n\}_{n=0}^{\infty}$, are calculated analytically or numerically giving $\{(H_0)_{nm}\}_{n,m=0}^{M-1}$ for some integer $M \geq N$. The $N\times N$ matrix $\tilde{V}$ is then added to the $N\times N$ submatrix of $H_0$ to give the total $N\times N$ Hamiltonian matrix $H$. The eigenvalues $\{\xi_n\}_{n=0}^{N-1}$ and corresponding normalized eigenvectors $\{\Gamma_{mn}\}_{m=0}^{N-1}$ of the matrix $H$ are calculated. Finally, the finite Green's function $g_{n,m}(E)$ is written as

$$g_{N-1,N-1}(E) = \langle \chi_{N-1} | [H-E]^{-1} | \chi_{N-1} \rangle = \sum_{n=0}^{N-1} \frac{\Gamma_{N-1,n}^2}{\xi_n - E} \tag{3.13}$$

3. We evaluate the set of eigenvalues $\{\varepsilon_k\}_{k=0}^{N-1}$, $\{\hat{\varepsilon}_k\}_{k=0}^{N-2}$, $\{\tilde{\varepsilon}_k\}_{k=0}^{N-2}$, $\{\bar{\varepsilon}_k\}_{k=0}^{N-3}$, $\{\varepsilon'_k\}_{k=0}^{M-1}$, and $\{\hat{\varepsilon}'_k\}_{k=0}^{M-2}$ associated with the matrices $\{(H_0)_{nm}\}_{n,m=0}^{N-1}$, $\{(H_0)_{nm}\}_{n,m=1}^{N-1}$, $\{(H_0)_{nm}\}_{n,m=0}^{N-2}$, $\{(H_0)_{nm}\}_{n,m=1}^{N-2}$, $\{(H_0)_{nm}\}_{n,m=0}^{M-1}$, and $\{(H_0)_{nm}\}_{n,m=1}^{M-1}$, respectively.
4. Using these eigenvalues, we obtain the following polynomial ratios:
   4.1 $[q_N(E)/p_N(E)]$ using the two sets of eigenvalues $\{\varepsilon_n\}_{n=0}^{N-1}$ and $\{\hat{\varepsilon}_n\}_{n=0}^{N-2}$.
   4.2 $[q_{N-1}(E)/p_{N-1}(E)]$ using the two sets of eigenvalues $\{\tilde{\varepsilon}_n\}_{n=0}^{N-2}$ and $\{\bar{\varepsilon}_n\}_{n=0}^{N-3}$.
   4.3 $[p_N(E)/p_{N-1}(E)]$ using the two sets of eigenvalues $\{\varepsilon_n\}_{n=0}^{N-1}$ and $\{\tilde{\varepsilon}_n\}_{n=0}^{N-2}$.
   4.4 $\tilde{G}_{00}(z)$ using the two sets of eigenvalues $\{\varepsilon'_n\}_{n=0}^{M-1}$ and $\{\hat{\varepsilon}'_n\}_{n=0}^{M-2}$.
5. The analytic continuation method is applied to $\tilde{G}_{00}(z)$ expression in (3.12) for a given choice of dimension $M$ and fit order $K$. The result is an approximation for the analytic function $G_{00}(z)$. This together with the polynomial ratios obtained in the previous step will be used to calculate the coefficients $T_{N-1}(E)$ and $R_N^{\pm}(E)$ as given by equations (3.8) and (3.9), respectively.

The above development assumes that the chosen basis is orthogonal. However, if it is not, then a modification in some of the formulas used above is necessary. This is carried out in the Appendix where the modified expressions are given primed equation numbers corresponding to the respective equations above.

## IV. EXAMPLES

We apply the numerical algorithm developed in the previous section to four examples with the same type of short-range potential, $\tilde{V}(r) \sim r^2 e^{-r}$, but for different reference Hamiltonians and bases. The third example is an extension of the scheme to multi-channel scattering. The fourth example is a relativistic Coulomb scattering where the calculation is done in the non-relativistic limit so that we can compare with known analytic non-relativistic J-matrix results. We implemented the numerical algorithm by using simple computer codes written in Mathcad®7. These codes are available upon request directly from the authors.

**Example (1): Coulomb scattering in the orthogonal Laguerre basis**

® Mathcad is a scientific programming package made by Mathsoft Corporation

This is the original J-matrix problem [4-6] that is usually solved in the following non-orthogonal Laguerre basis in which the matrix representation of $H_0$ is tridiagonal:

$$\chi_n(r) = \sqrt{\frac{\lambda\,\Gamma(n+1)}{\Gamma(2l+n+2)}}\,(\lambda\,r)^{l+1} e^{-\lambda\,r/2} L_n^{2l+1}(\lambda\,r) \qquad (4.1)$$

where $\lambda$ is the basis scale parameter and $L_n^{\nu}(x)$ is the generalized Laguerre polynomial defined in accordance with the notation and convention of reference [23]. However, our choice of basis will be the following orthonormal Laguerre set

$$\chi_n(r) = \sqrt{\frac{\lambda\,\Gamma(n+1)}{\Gamma(2l+n+3)}}\,(\lambda\,r)^{l+1} e^{-\lambda\,r/2} L_n^{2l+2}(\lambda\,r) \qquad (4.2)$$

In this basis the matrix representation of $H_0$ is not tridiagonal but rather takes the following form

$$\left(H_0\right)_{nm} = \frac{\lambda^2}{4}\left[\sqrt{\frac{\Gamma(n_>+1)\Gamma(2l+n_<+3)}{\Gamma(n_<+1)\Gamma(2l+n_>+3)}}\left(1+\frac{2n_<}{2l+3}+\frac{2Z/\lambda}{l+1}-\frac{1}{2}\delta_{nm}\right)\right] \qquad (4.3)$$

where $n_>$ ($n_<$) is the larger (smaller) of the two integers $n$ and $m$. The numerical scheme is applied to this problem with the short range potential $\tilde{V}(r) = 7.5r^2e^{-r}$ and with the following parameters:

$$N = 30,\ M = 40,\ K = 20,\ \lambda = 2,\ Z = 1,\ l = 0 \qquad (4.4)$$

Figure (1) is a plot of $\left|1 - S(E)\right|$ vs. $E$ (in atomic units) resulting from the numerical scheme while Figure (2) shows the original J-matrix analytic result $\left|1 - S^{AN}(E)\right|$. It is quite clear that the agreement in the given energy range is excellent.

**Example (2): Coulomb scattering in the orthogonal oscillator basis**

Traditionally, the orthonormal oscillator basis

$$\chi_n(r) = \sqrt{\frac{2\lambda\,\Gamma(n+1)}{\Gamma(l+n+3/2)}}\,(\lambda r)^{l+1} e^{-\lambda^2 r^2/2} L_n^{l+1/2}(\lambda^2 r^2) \qquad (4.5)$$

is not associated with the J-matrix Coulomb problem. This is due to the fact that the representation of the Coulomb potential *Z/r* in this basis is not tridiagonal. Historically, this basis was used with a reference Hamiltonian that is made up of the kinetic energy only without any other analytic potential term. It was shown, however, that including the oscillator potential $\frac{1}{2}\tau r^2$ in $H_0$ preserves the tridiagonal structure in this oscillator basis and maintains compatibility with the J-matrix formalism [14,15]. Moreover, it is worthwhile noting that this type of basis is frequently used in nuclear physics, which makes the numerical J-matrix scheme proposed here a good candidate for nuclear scattering calculations. Anyhow, for the present problem, the matrix representation of the kinetic energy part of $H_0$ is tridiagonal and well known [5]. We choose to calculate the matrix elements of the Coulomb potential part of $H_0$ in this basis using Gauss quadrature integral approximation [24]. The numerical scheme is applied with the same short range potential $\tilde{V}(r) = 7.5r^2e^{-r}$. We took the following parameter values:

$$N = 30,\ M = 40,\ K = 20,\ \lambda = 1,\ Z = 1,\ l = 1 \qquad (4.6)$$

The result is shown in Figure (3). It is to be compared with the analytic result calculated in the Laguerre basis and shown in Figure (4). The agreement is very good.

**Example (3): Two-channel scattering in the orthogonal oscillator basis**

The numerical scheme is easily extendable to multi-channel scattering [9,10]. We consider a two-channel scattering using the orthogonal oscillator basis for a reference Hamiltonian having only the kinetic energy term and with threshold energies (in atomic units):

$$E=\begin{pmatrix}0.0\\0.1\end{pmatrix} \tag{4.7}$$

The two channels are coupled by the following short-range potential

$$\tilde{V}(r)=\begin{pmatrix}-1.0 & -7.5\\-7.5 & 7.5\end{pmatrix}r^2e^{-r} \tag{4.8}$$

This potential has been considered by Noro and Taylor [25] and by Mandelshtam *et al* [26]. The calculation is carried out for S-wave scattering with the following two-cannel parameters

$$N=\begin{pmatrix}20\\25\end{pmatrix}, M=\begin{pmatrix}30\\35\end{pmatrix}, K=\begin{pmatrix}30\\30\end{pmatrix}, \lambda=\begin{pmatrix}2\\2\end{pmatrix} \tag{4.9}$$

The results of the numerical scheme are shown graphically in Figures (5.1) and (5.2) for the two-coupled channels. They agree very well with the corresponding analytic results in Figures (6.1) and (6.2). In the graphs $S_{nm}(E)$ stands for the scattering matrix that couples channel $n$ and $m$.

**Example (4): Relativistic Coulomb scattering in the relativistic Laguerre basis**

The two-component radial Dirac Hamiltonian for a charged spinor in the Coulomb field, $Z/r$, is [12]

$$H_0=\begin{pmatrix}-\dfrac{\gamma}{\kappa} & \alpha\left(\dfrac{Z}{\kappa}-\dfrac{\gamma}{r}-\dfrac{d}{dr}\right)\\ \alpha\left(\dfrac{Z}{\kappa}-\dfrac{\gamma}{r}+\dfrac{d}{dr}\right) & \dfrac{\gamma}{\kappa}+2\dfrac{\alpha^2 Z}{r}\end{pmatrix} \tag{4.10}$$

where $\alpha$ is the fine structure constant and $\kappa$ is the spin-orbit coupling parameter defined by

$$\kappa=\pm(j+½) \text{ for } l=j\pm ½ \tag{4.11}$$

$j$ is the total angular momentum quantum number and $\gamma=\sqrt{\kappa^2-(\alpha Z)^2}$. The spinor wave function basis $\{\chi_n(r)\}_{n=0}^{\infty}$ has the following two components [12]

$$\vartheta_n(r)=a_n(\lambda r)^{\gamma+1}e^{-\lambda r/2}L_n^{2\gamma+1}(\lambda r) \tag{4.12}$$

$$\phi_n(r)=\frac{\lambda C}{2}a_n(\lambda r)^{\gamma}e^{-\lambda r/2}\left[(2\gamma+n+1)L_n^{2\gamma}(\lambda r)+(n+1)L_{n+1}^{2\gamma}(\lambda r)\right] \tag{4.13}$$

where $\lambda$ is the basis scale parameters, $C$ is the small component strength parameter, and $a_n$ is the normalization constant

$$a_n=\sqrt{\frac{\lambda\Gamma(n+1)}{2\Gamma(2\gamma+n+2)}} \tag{4.14}$$

This is also an example for the case in which the chosen basis is not orthogonal and, thus, one is required to use the expressions given in the Appendix. The representation of $H_0$ in this basis is tridiagonal [12], however, the resulting recursion relation is not simple to solve analytically. Figure (7) shows the results of the numerical scheme calculations in the non-relativistic limit as a plot of $|1-S(\varepsilon)|$ vs. $E$ (in atomic units), where $\varepsilon$ is the relativistic energy. Figure (8) is a plot of $|1-S_{NR}^{AN}(E)|$ vs. $E$ (in atomic

units) that is obtained from the non-relativistic analytic J-matrix calculation for the same problem with the same sub-parameters. The energy variables are related by $\varepsilon \cong \gamma/\kappa + \alpha^2 E$. The parameters were taken in the non-relativistic limit with the following values:

$$\alpha = 10^{-3}, Z = 1, |\kappa| = l = 2, \lambda = 2, C = -\alpha/2, N = M = 30, K = 25 \qquad (4.15)$$

It is clear that the agreement is excellent. The relativistic effects have been calculated in reference [12] using this J-matrix numerical scheme.

## ACKNOWLEDGMENTS

The financial support by KFUPM is greatly acknowledged by the third author (MSA).

## APPENDIX: CALCULATION IN NON-ORTHOGONAL BASES

In the case of non-orthogonal bases, some of the formulas used in the numerical scheme of section III have to be modified. The modified expressions are given below with primed equation numbers that correspond to the respective equations in section III. We start by defining the quantities used in the modified formulas.

Let $\{\chi_n\}_{n=0}^{\infty}$ be the basis of the $L^2$ space whose conjugate is spanned by $\{\overline{\chi}_n\}_{n=0}^{\infty}$. That is $\langle \overline{\chi}_n | \chi_m \rangle = \delta_{nm}$ whereas the basis-overlap matrix $\langle \chi_n | \chi_m \rangle \equiv \theta_{nm} \neq \delta_{nm}$. It is assumed that all matrix representations of operators are given in the basis $\{\chi_n\}$. In subsequent expressions we use the following defined quantities:

- $\theta$, $\hat{\theta}$, $\tilde{\theta}$, $\overline{\theta}$, $\theta'$, and $\hat{\theta}'$ stand for the basis-overlap matrices whose elements are $\{\theta_{ij}\}_{i,j=0}^{N-1}$, $\{\theta_{ij}\}_{i,j=1}^{N-1}$, $\{\theta_{ij}\}_{i,j=0}^{N-2}$, $\{\theta_{ij}\}_{i,j=1}^{N-2}$, $\{\theta_{ij}\}_{i,j=0}^{M-1}$, and $\{\theta_{ij}\}_{i,j=1}^{M-1}$, respectively.
- $H_0, \hat{H}_0, \tilde{H}_0, \overline{H}_0, H_0', \hat{H}_0'$ are similarly and respectively defined.
- $|\theta|$, $|\hat{\theta}|$, $|\tilde{\theta}|$, $|\overline{\theta}|$, $|\theta'|$, and $|\hat{\theta}'|$ stand for the determinants of their corresponding matrices.
- $\{\varepsilon_k\}_{k=0}^{N-1}$, $\{\hat{\varepsilon}_k\}_{k=0}^{N-2}$, $\{\tilde{\varepsilon}_k\}_{k=0}^{N-2}$, $\{\overline{\varepsilon}_k\}_{k=0}^{N-3}$, $\{\varepsilon_k'\}_{k=0}^{M-1}$, and $\{\hat{\varepsilon}_k'\}_{k=0}^{M-2}$ is the set of generalized eigenvalues for $H_0$, $\hat{H}_0$, $\tilde{H}_0$, $\overline{H}_0$, $H_0'$, and $\hat{H}_0'$ with respect to the overlap matrix $\theta$, $\hat{\theta}$, $\tilde{\theta}$, $\overline{\theta}$, $\theta'$, and $\hat{\theta}'$, respectively.
- $\xi_n$ is the generalized eigenvalue of $H$ with respect to the overlap matrix $\theta$ and $\{\Gamma_{m,n}\}_{m=0}^{N-1}$ is the associated normalized eigenvector satisfying the generalized eigenvalue equation: $H|\Gamma_n\rangle = \xi_n \theta |\Gamma_n\rangle$.
- $\sigma_n$ is defined by the diagonalization equation $(\Gamma^{T}\theta\ \Gamma)_{nm} = \sigma_n \delta_{nm}$.
- $\tilde{\xi}_n$ is the generalized eigenvalue of $\tilde{H}$ (the submatrix of $H$ whose elements are $\{H_{nm}\}_{n,m=0}^{N-2}$) with respect to the overlap matrix $\tilde{\theta}$.

With these definitions, we rewrite the relevant expressions given in section III for the case of non-orthogonal basis as follows:

$$\frac{q_N(z)}{p_N(z)} = \frac{|\hat{\theta}|}{|\theta|} \frac{\prod_{k=0}^{N-2}(z-\hat{\varepsilon}_k)}{\prod_{k=0}^{N-1}(z-\varepsilon_k)} \tag{3.10'}$$

$$\frac{q_{N-1}(z)}{p_{N-1}(z)} = \frac{|\bar{\theta}|}{|\tilde{\theta}|} \frac{\prod_{k=0}^{N-3}(z-\bar{\varepsilon}_k)}{\prod_{k=0}^{N-2}(z-\tilde{\varepsilon}_k)} \tag{3.10''}$$

$$J_{N-1,N}(z)\frac{p_N(z)}{p_{N-1}(z)} = \frac{|\theta|}{|\tilde{\theta}|} \frac{\prod_{k=0}^{N-1}(z-\varepsilon_k)}{\prod_{k=0}^{N-2}(z-\tilde{\varepsilon}_k)} \tag{3.11'}$$

$$\tilde{G}_{00}(z) = -\frac{|\hat{\theta}'|}{|\theta'|} \frac{\prod_{k=0}^{M-2}(z-\hat{\varepsilon}'_k)}{\prod_{k=0}^{M-1}(z-\varepsilon'_k)} \qquad ; M \geq N \tag{3.12'}$$

$$g_{N-1,N-1}(E) = \langle \bar{\chi}_{N-1} | [H-E]^{-1} | \bar{\chi}_{N-1} \rangle = \sum_{n=0}^{N-1} \frac{1}{\sigma_n} \frac{\Gamma^2_{N-1,n}}{\xi_n - E}$$

$$= \frac{|\tilde{\theta}|}{|\theta|} \frac{\prod_{m=0}^{N-2}(\tilde{\xi}_m - E)}{\prod_{n=0}^{N-1}(\xi_n - E)} \tag{3.13'}$$

## FIGURES CAPTION:

**FIG. 1:** A plot of $\left|1-S(E)\right|$ vs. $E$ (in atomic units) resulting from the numerical scheme in the unconventional (in the J-matrix sense) orthogonal Laguerre basis for the Coulomb scattering problem of Example (1). The short-range potential is $\tilde{V}(r)=7.5r^2e^{-r}$ and the parameter values are: $N = 30$, $M = 40$, $K = 20$, $\lambda = 2$, $Z = 1$, $l = 0$.

**FIG. 2:** Shows the original J-matrix analytic result of $\left|1-S^{AN}(E)\right|$ for the same problem of Example (1) with the parameters: $N = 30$, $\lambda = 2$, $Z = 1$, $l = 0$.

**FIG. 3:** A plot of $\left|1-S(E)\right|$ vs. $E$ (in atomic units) resulting from the numerical scheme in the non-traditional oscillator basis for the same Coulomb scattering problem. The parameters were assigned the following values: $N = 30$, $M = 40$, $K = 20$, $\lambda = 1$, $Z = 1$, $l =$ 1.

**FIG. 4:** The J-matrix analytic result of $\left|1-S^{AN}(E)\right|$ for the same Coulomb scattering problem of Example (2) in the original Laguerre basis and with the same physical parameters.

**FIG. 5:** Shows graphically the result obtained by the multi-channel extension of the numerical scheme applied to the two-channel problem of Example (3) with the short-range potential given by Eq. (4.8). Calculation is carried out in the orthogonal oscillator basis with channel parameters given in (4.9). $S_{nm}(E)$ is the scattering matrix that couples channel $n$ and $m$.

**FIG. 6:** shows the analytic result using the original multi-channel J-matrix for the same two-channel problem of Example (3).

**FIG. 7:** Shows the result obtained by the relativistic extension of the numerical scheme, however, in the non-relativistic limit as a plot of $\left|1-S(\varepsilon)\right|$ vs. $E$ (in atomic units), where $\varepsilon \cong \gamma/\kappa + \alpha^2 E$. The parameters were taken in the non-relativistic limit as shown in (4.15).

**FIG. 8:** A plot of $\left|1-S_{NR}^{AN}(E)\right|$ vs. $E$ (in atomic units) that is obtained from the non-relativistic J-matrix analytic method for the problem of Example (4).

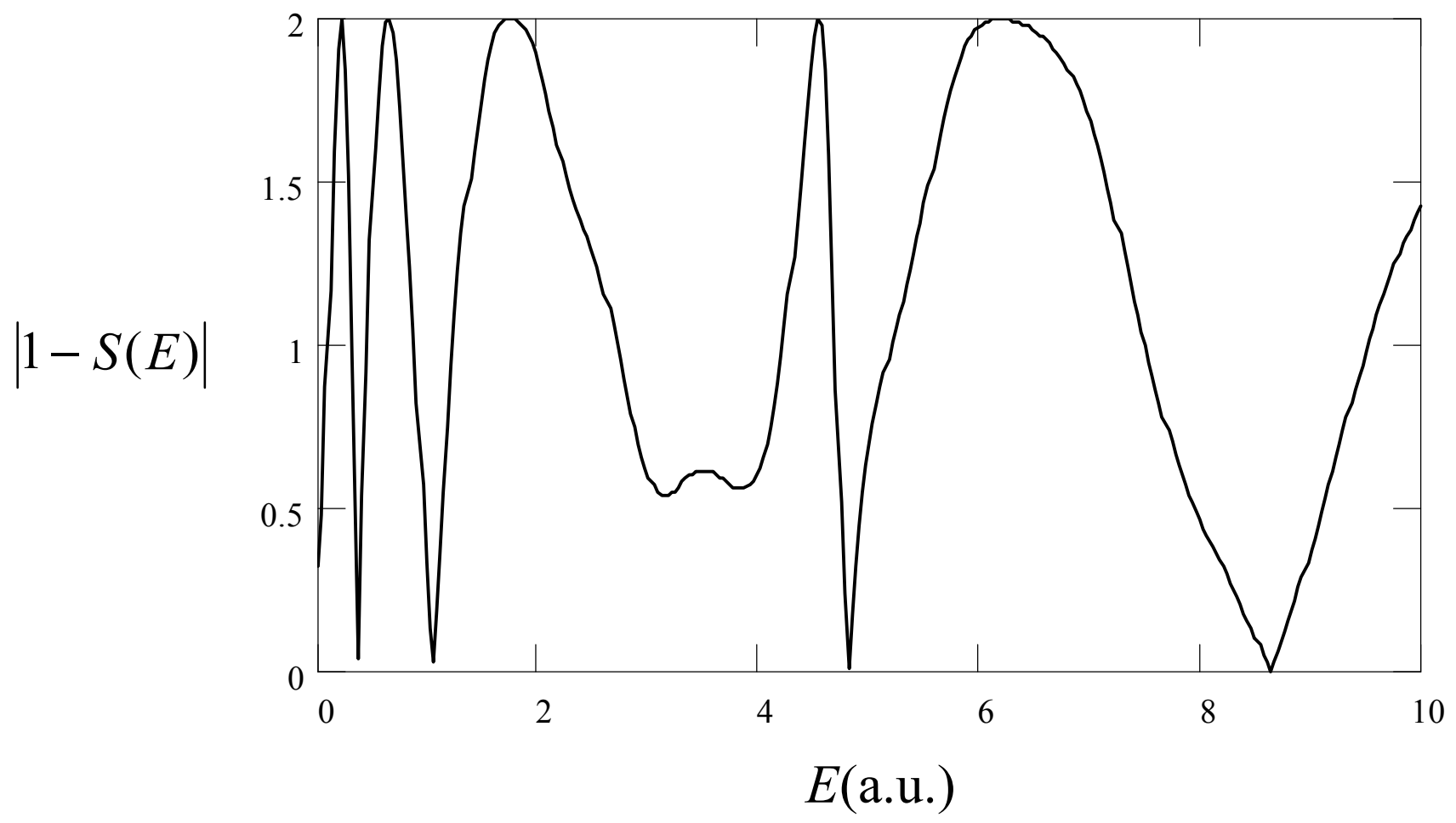

Figure (1)

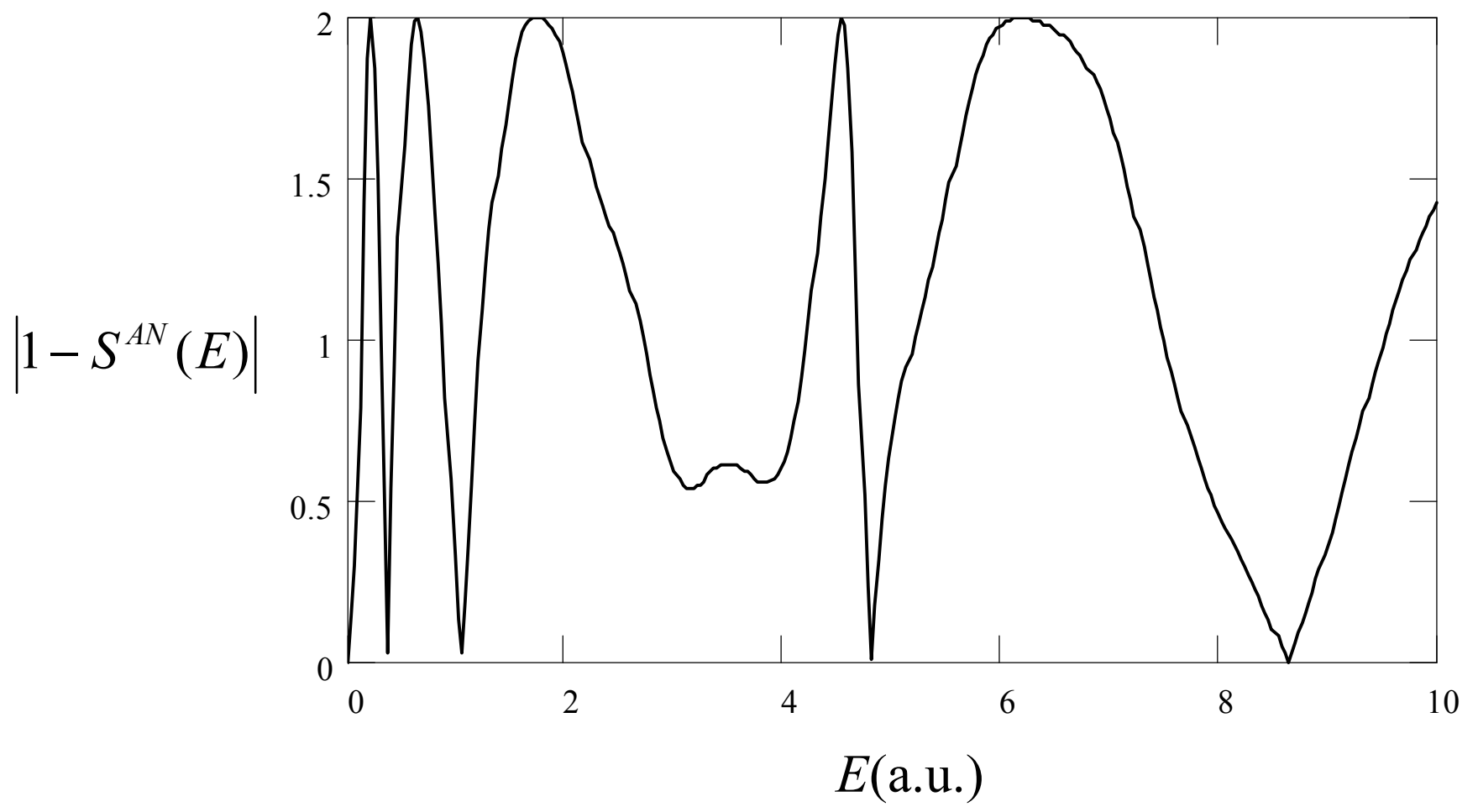

Figure (2)

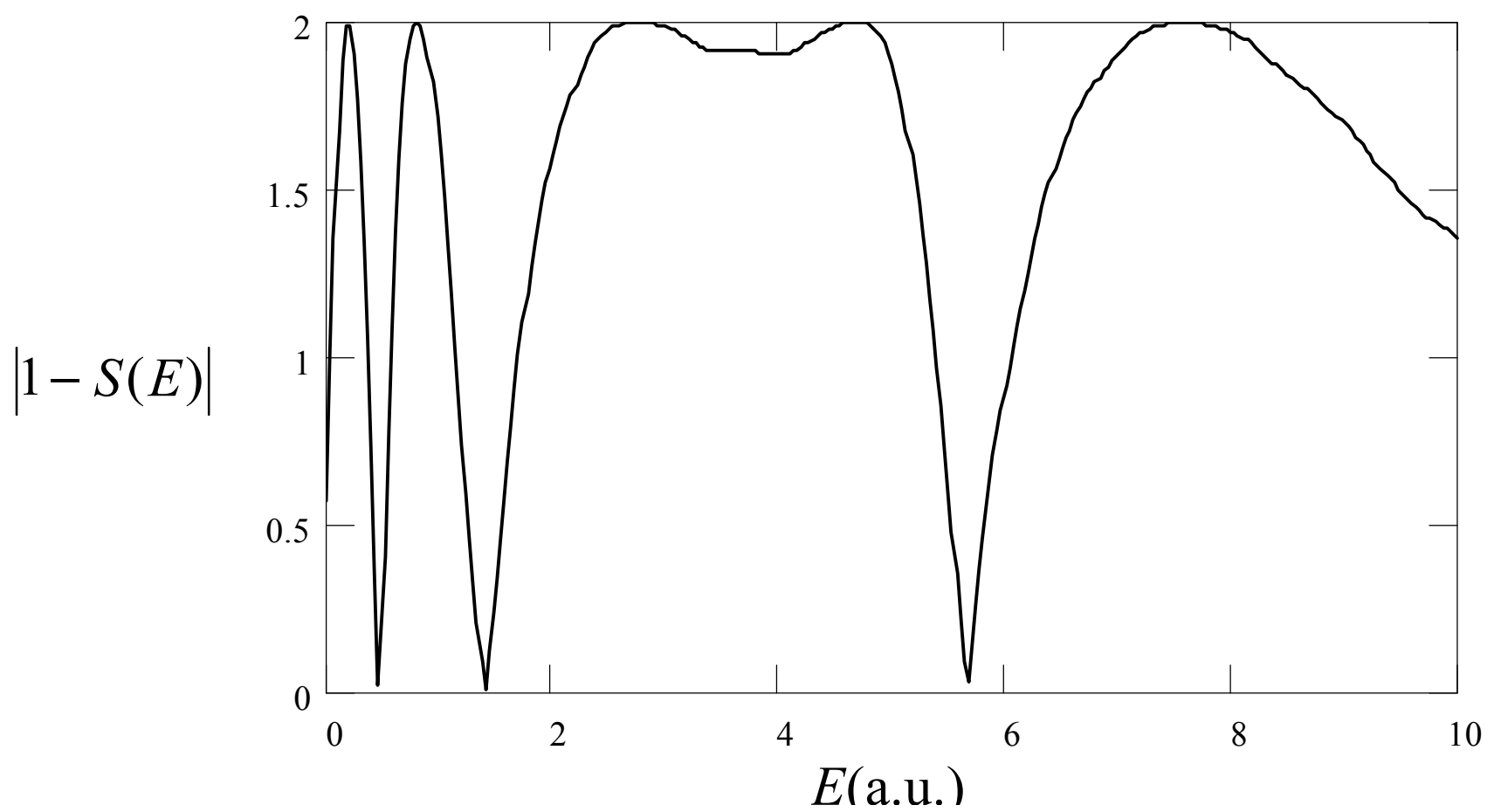

Figure (3)

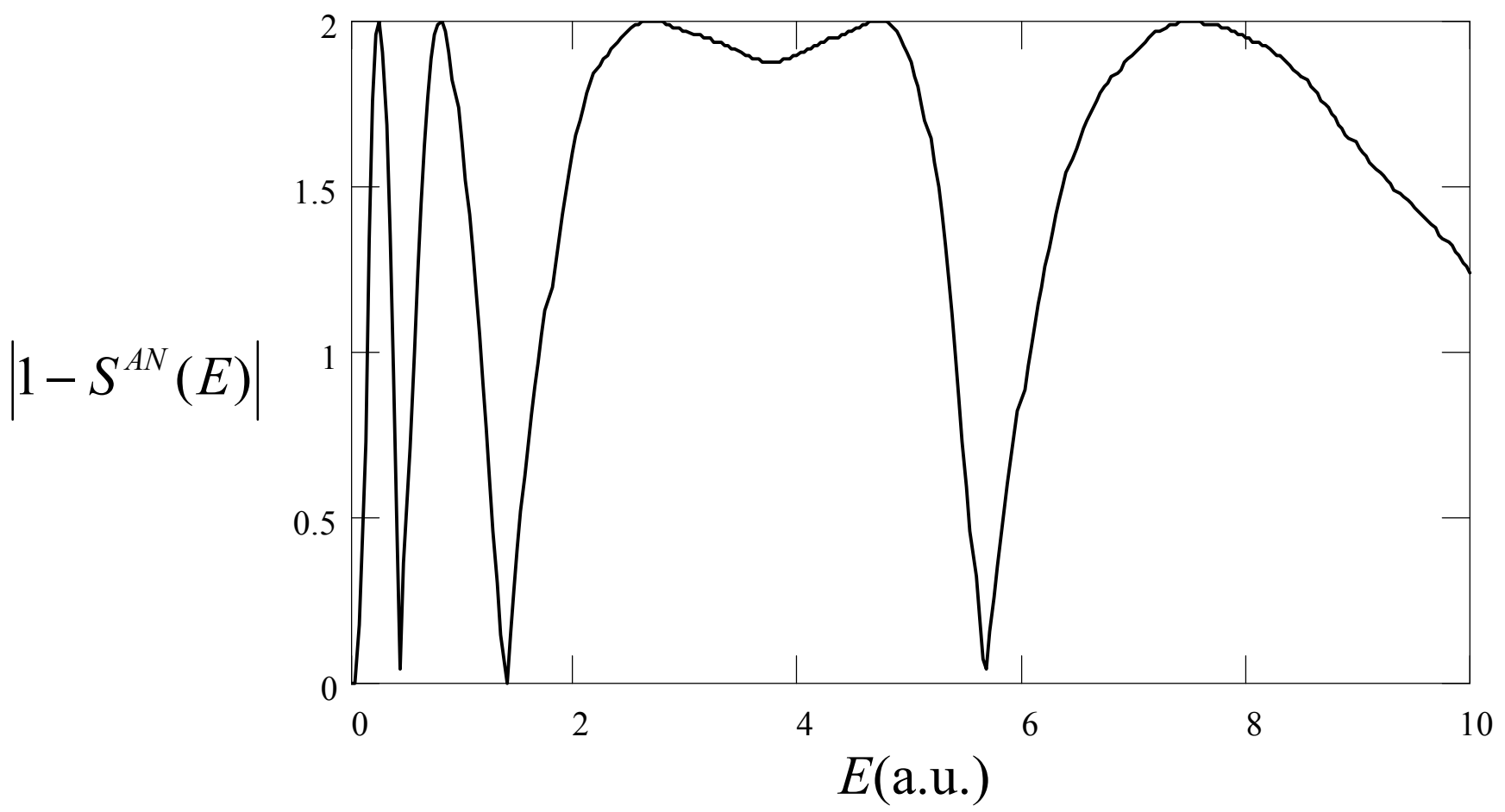

Figure (4)

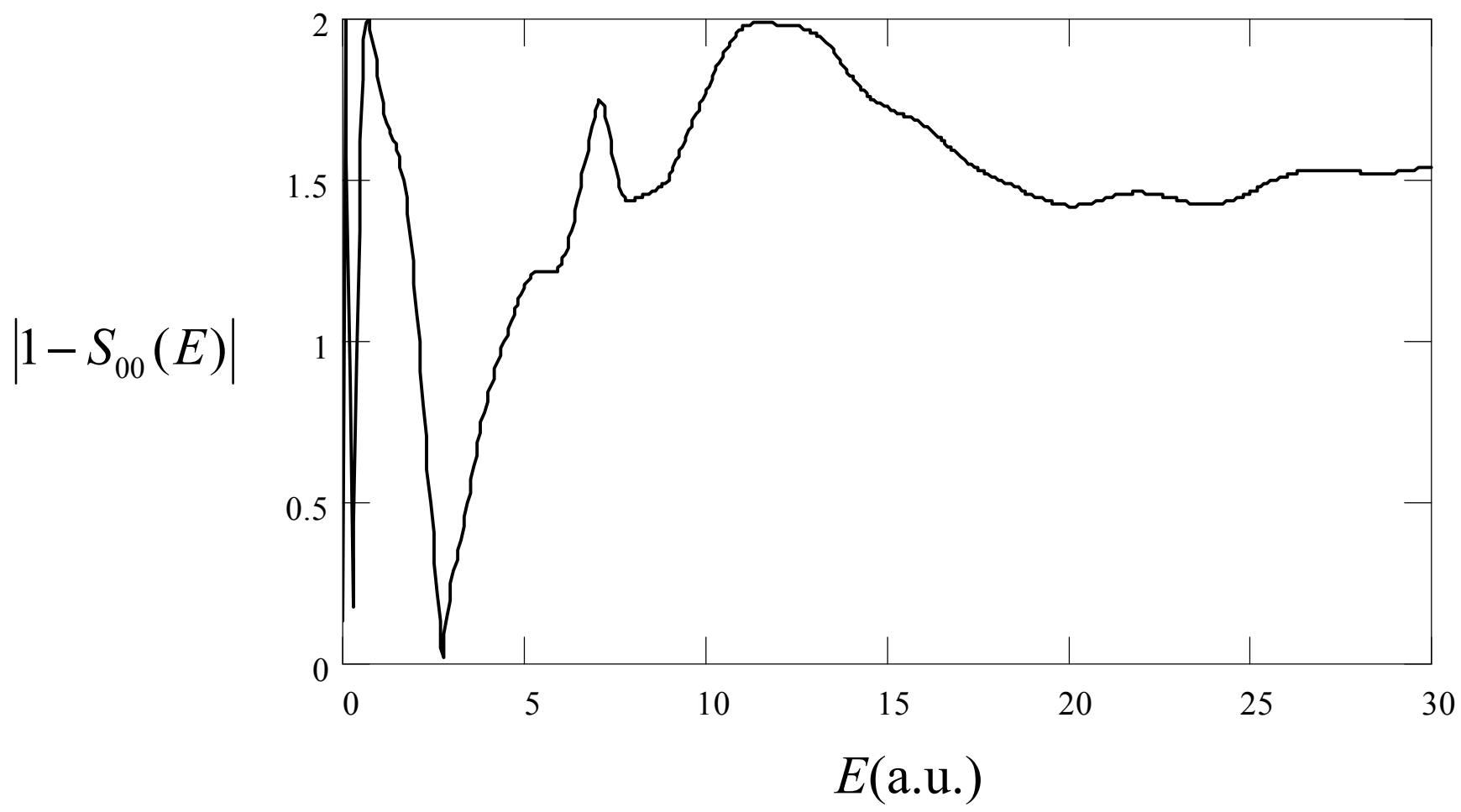

Figure (5.1)

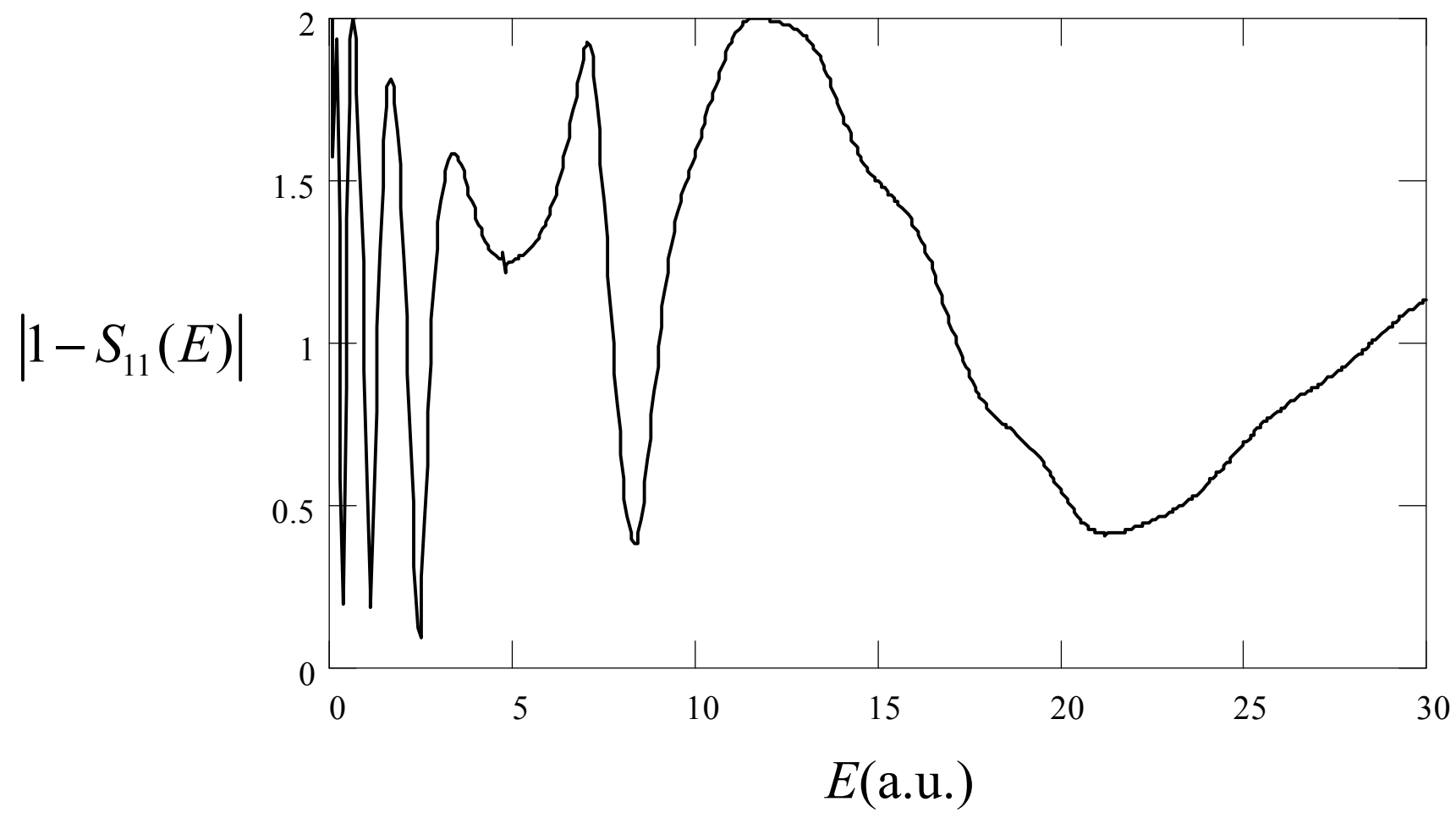

Figure (5.2)

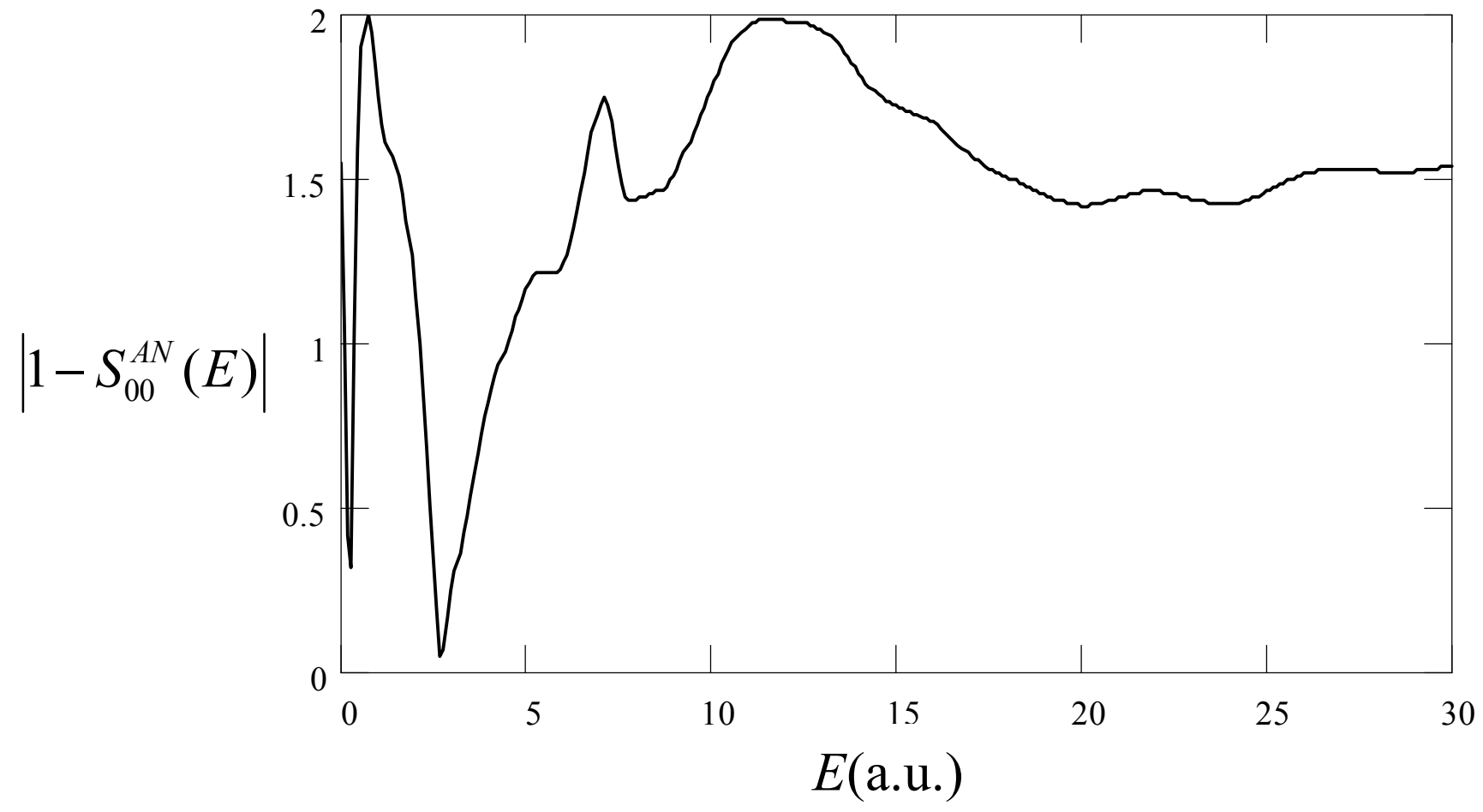

Figure (6.1)

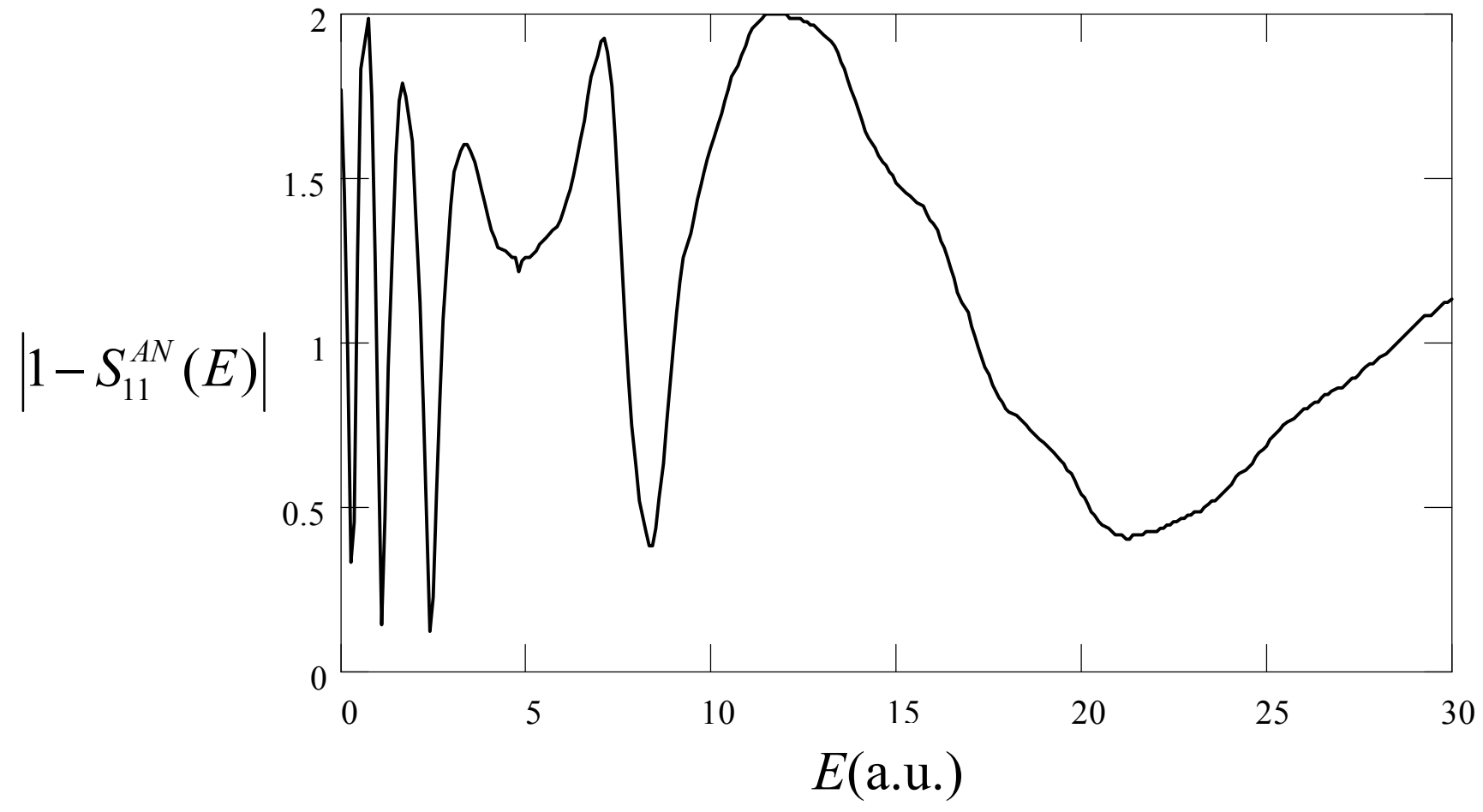

Figure (6.2)

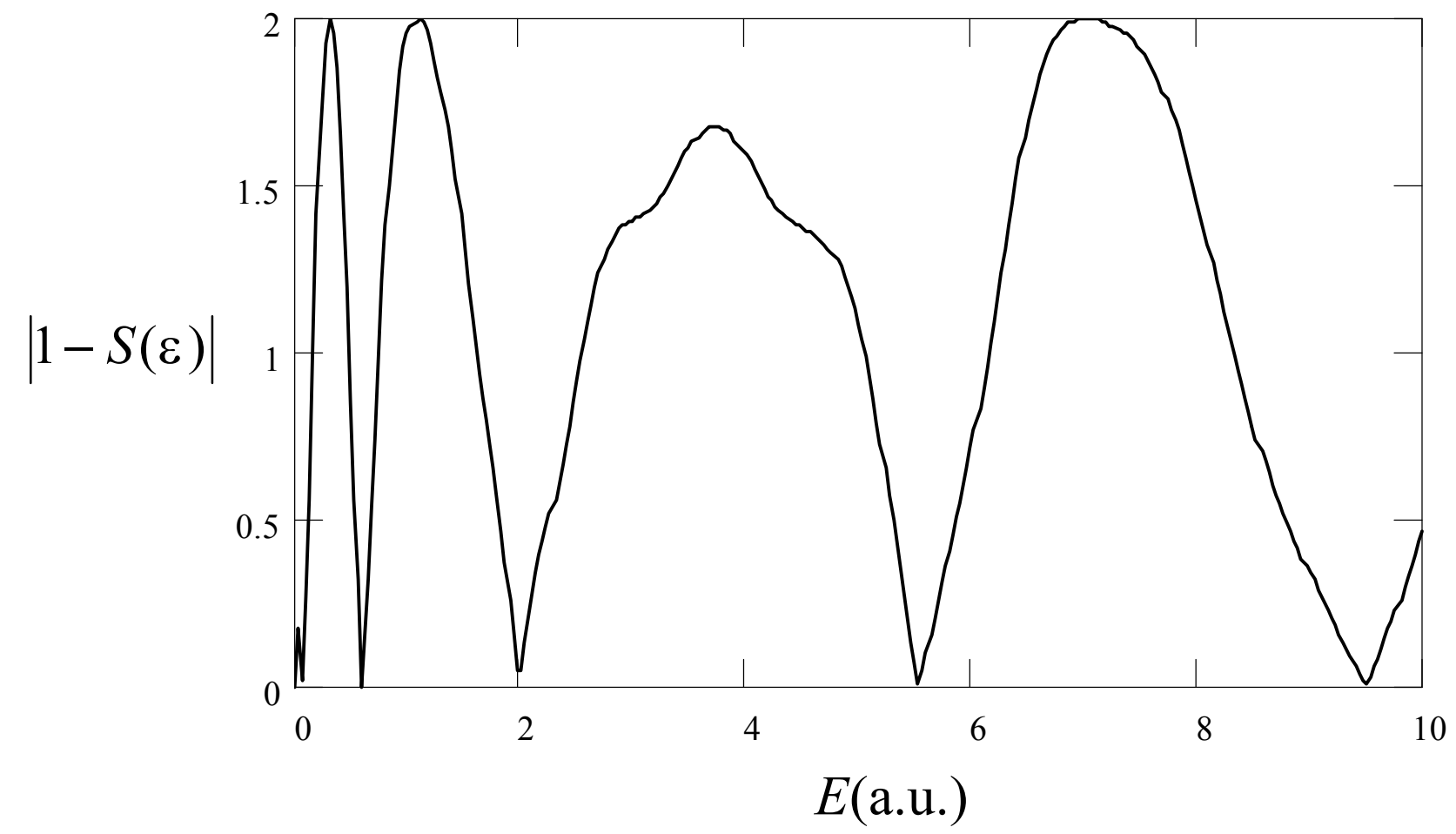

Figure (7)

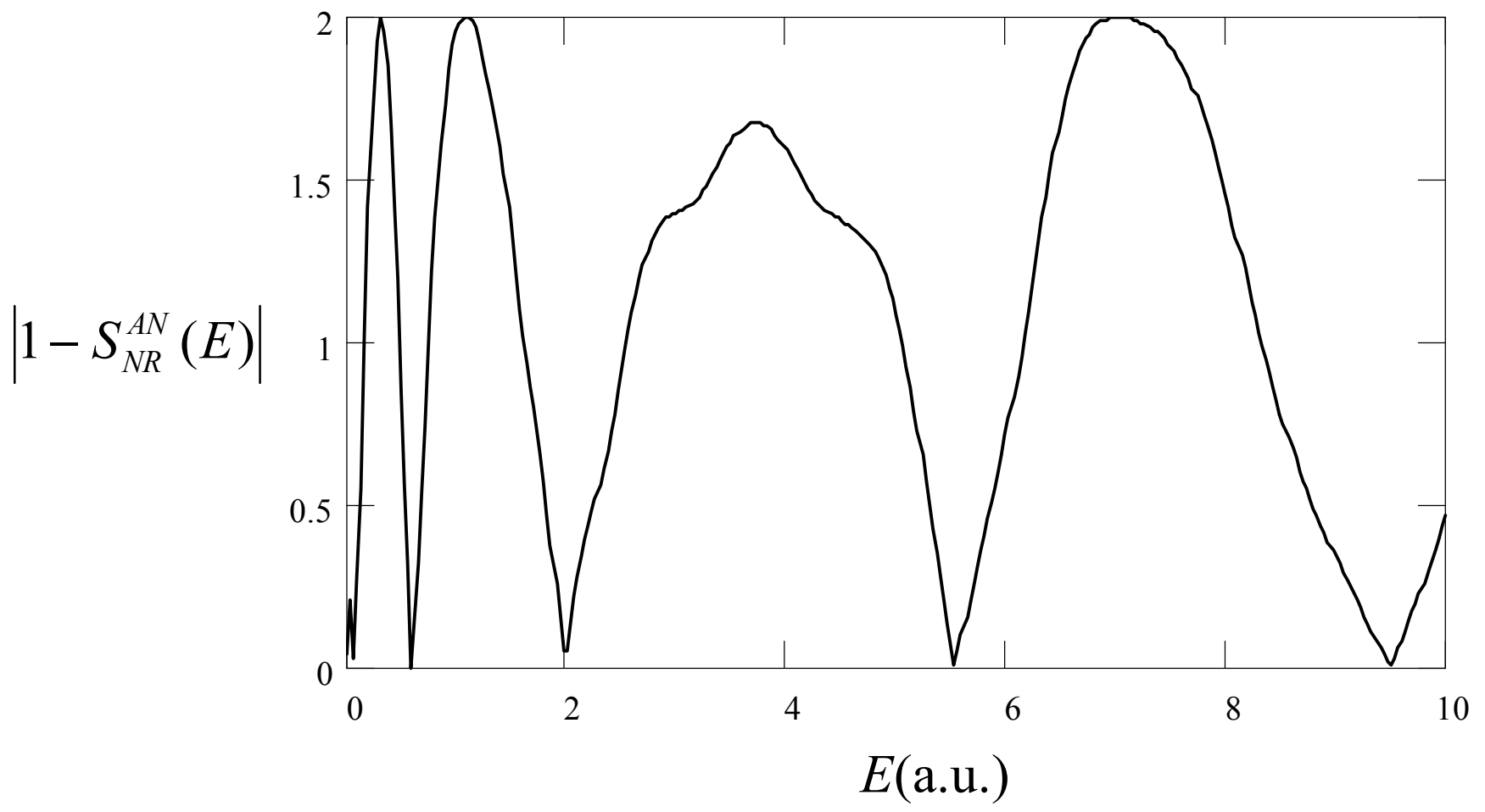

Figure (8)